\documentclass[10pt,prl,aps,twocolumn,superscriptaddress,floatfix,showpacs]{revtex4}

\pdfoutput=1

\usepackage{graphicx}
\usepackage{amssymb}
\usepackage{bm}
\usepackage{bbold}
\usepackage[toc,page]{appendix}

\begin{document}

\title{Chiral topological spin liquids with projected entangled pair states}

\author{Didier Poilblanc}
\affiliation{Laboratoire de Physique Th\'eorique, C.N.R.S. and Universit\'e de Toulouse, 31062 Toulouse, France}

\author{J.~Ignacio \surname{Cirac}}
\affiliation{Max-Planck-Institut f{\"{u}}r Quantenoptik,
Hans-Kopfermann-Str.\ 1, D-85748 Garching, Germany}

\author{Norbert \surname{Schuch}}
\affiliation{Institut f\"ur Quanteninformation, RWTH Aachen, D-52056 Aachen, Germany}

\date{\today}

\begin{abstract}
  Topological chiral phases are ubiquitous in the physics of the Fractional Quantum Hall Effect.
  Non-chiral topological spin liquids are also well known. Here, using the framework of  projected entangled pair states (PEPS), we construct a family of chiral spin liquids on the square lattice
  which are  generalized spin-1/2 Resonating Valence Bond (RVB) states
obtained from deformed local tensors with $d+i\, d$ symmetry.  
On a cylinder, we construct four topological sectors with even or odd number of spinons on the boundary and even or odd number of ($\mathbb{Z}_2$) fluxes penetrating the cylinder which, we  argue,
remain orthogonal in the limit of infinite perimeter. The analysis of the transfer matrix provides evidence 
 of short-range (long-range) triplet (singlet) correlations as for the critical (non-chiral) RVB state.
The Entanglement Spectrum exhibits chiral edge modes, which we confront to predictions of
Conformal Field Theory, and the corresponding Entanglement Hamiltonian is shown to be long ranged.   
\end{abstract}
\maketitle

{\it Introduction} -- 
The study of topological states of matter is the focus of research of a
multidisciplinary scientific community. On the one hand, they give rise to very
exotic behavior associated to the topological order~\cite{TO}, something which has no analog in standard Condensed Matter
Physics.
On the other, it provides an alternative and promising approach to fault
tolerant quantum computing~\cite{TC}. Furthermore, experiments both with
new materials~\cite{nayakRMP} and cold atoms~\cite{ColdAtoms} are now at the position of engineering some
of those states, and thus provide a laboratory to verify
the theoretical findings.

Two of the main theoretical challenges in the study of topological matter are the
identification of simple models and descriptions which can lend us a
complete understanding of their exotic properties, as well as the derivation
of parent Hamiltonians, so that we get a hint of how to create the states, or
under which conditions we can find them at low temperature. A possible
approach to attack both challenges is through the use of tensor networks, and
in particular of Projected Entangled Pair States (PEPS)~\cite{frank1,cirac}. They provide 
a simple local description of many body states from which their physical properties can be determined, and
they imply the existence of a parent local Hamiltonian.

So far, most of the PEPS studies have concentrated on non-chiral topologically ordered
states, like the toric code~\cite{cirac,TC}, double and Levin-Wen  models~\cite{LevinWen}, or 
the short-range spin--1/2 RVB states~\cite{frank2,norbert,didier,wang}. 
The latter is
specially appealing as it respects su(2) symmetry and is translationally invariant, something
which naturally appears in many materials~\cite{RVB0}. Furthermore, in the Kagome lattice
it corresponds to a $\mathbb{Z}_2$ spin 
liquid~\cite{norbert,didier}, with very well known anyonic
excitations.

Another class of topological ordered matter corresponds to the Fractional Quantum Hall State (FQHS) exhibiting
protected chiral edge modes~\cite{wen} that are fully characterized by a chiral CFT~\cite{CFT}.
For example, the edge states of the bosonic Laughlin~\cite{laughlin}, Moore-Read~\cite{moore} and
Read-Rezayi~\cite{read} states can be described by a chiral $SU(2)_k$ CFT of the
Wess-Zumino-Witten (WZW) model~\cite{CFT}, where k = 1; 2 and 3 respectively. The edge modes
are protected by the long-range topological order of the bulk which hosts
quasiparticles (or anyons) with Abelian ($k = 1$) and non-Abelian ($k > 1$) fractional statistics.
Recently, an emergent topological chiral spin liquid
(CSL), similar to the one proposed by Kalmeyer and
Laughlin (KL)~\cite{KL,WWZ}, has been discovered in the Kagome
quantum antiferromagnet with broken time-reversal (TR) symmetry~\cite{bauer} as well as in the square lattice~\cite{nielsen},
realizing the spin analogue of the bosonic Laughlin state. 
A FQHS was also found in a TR invariant Kagome Heisenberg model~\cite{sheng}.
Whether PEPS can be derived for gapped
chiral phases of interacting systems remains a major and
subtle issue debated in recent papers~\cite{dubail,wahl}. Nevertheless,
a critical chiral PEPS with topological order has
been constructed by Gutzwiller projecting two copies of
chiral states of free fermions~\cite{shuo}. Using the PEPS bulk-edge
correspondence, the edge CFT could be identified as $SO(2)_1$ but
with an infinite correlation length.

In this paper we show that by deforming the tensors corresponding to
the RVB spin--1/2 state, implementing $d+i\, d$ symmetry, one can break the TR symmetry
but keeping the su(2) and other symmetries of the state,
and obtain a chiral topological state. We analyze the topological degeneracies, 
the entanglement spectrum and the entropy via the bulk-boundary correspondence.
The chiral edge modes bear some features of CFT although a precise 
identification to simple known CFT was not possible.
We work on a square lattice, and find that, as the RVB state~\cite{albuquerque,tang}, the family
we construct has infinite correlation length and, hence, moves along a critical phase of a local Hamiltonian.

\begin{figure}
\begin{center}
\includegraphics[width=8cm,angle=0]{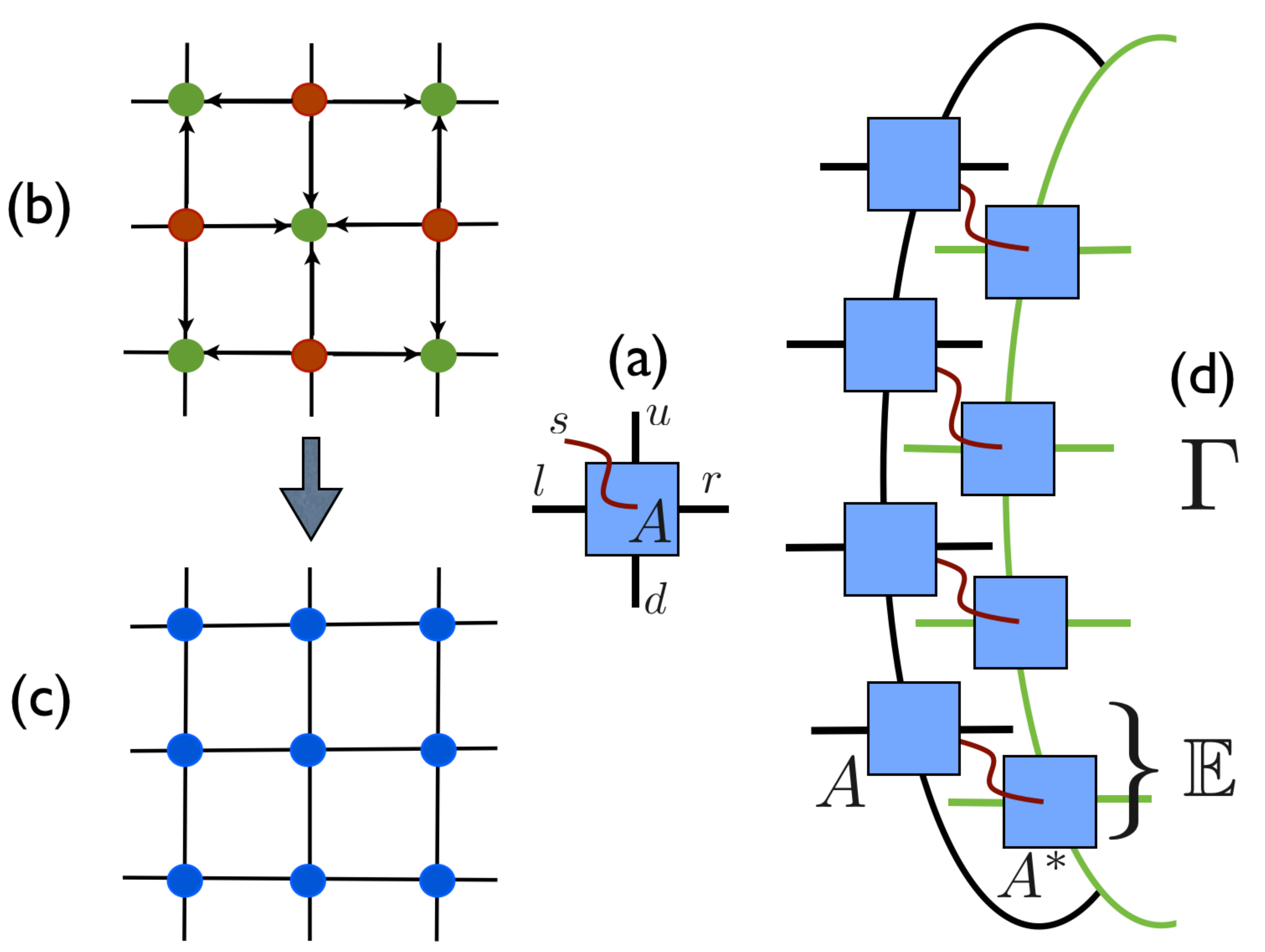}
\caption{(a) Local tensor $A$.
(b) Square lattice with (nearest-neighbor) singlet bonds oriented from sublattice A to sublattice B. (c) Under a 180-degrees spin rotation on 
e.g. the B sublattice, oriented singlets are transformed into symmetric 
$|\uparrow\uparrow\rangle+|\downarrow\downarrow\rangle$ maximally entangled pair states. (d) 
Transfer matrix on a periodic ring connecting the (non-contracted) BK virtual indices on the left of the ring to the ones on its right.
  }
\label{Fig:lattice}
\end{center}
\vskip -0.5cm
\end{figure}

{\it PEPS framework} -- 
We consider a two-dimensional (2d) bipartite square lattice where
a rank-5 tensor $A^s_{lurd}$ of dimension $dD^4$ is assigned to each site. Here, the index $s=0,1$ stands for the local physical spin-1/2 degrees of freedom (of dimension $d=2$) and the subscript indices $l,u,r,d$ label the virtual states on the four bonds as shown in Fig.~\ref{Fig:lattice}(a).
Basically, we can write a general spin-1/2 ansatz wave function on a periodic manifold of $M$ sites as $|\Psi\rangle = \sum_{s_1,s_2\dots i_M}c_{s_1,s_2\dots s_M}|s_1,s_2\dots s_M\rangle$, and $c_{s_1,s_2\dots s_M}={\rm Contract}[A^{s_1}A^{s_2}\dots A^{s_M}]$ where all tensors 
share the same bond variables with their 4 neighbors and ``Contract'' means that one sums up over all bond variables of the 2d tensor 
network. 
In order to construct a generalized spin-singlet RVB wavefunction, we assume that the virtual states belong to the $1/2\oplus 0$ spin representation of dimension $D=3$ (the virtual states $|\uparrow\rangle$, $|\downarrow\rangle$ and $|0\rangle$ carry 
$S_z=1/2,-1/2$ and $0$, respectively)~\cite{norbert,didier}. Since in each configuration of the RVB state, exactly one 
bond singlet $|\uparrow\downarrow\rangle-|\downarrow\uparrow\rangle$ 
is connected to each lattice site, the precise sign structure of the RVB
state depends on the orientations of the latter~: hereafter singlets 
are oriented from one sublattice (A) to the other sublattice (B) as shown in Fig.~\ref{Fig:lattice}(b).  
After  a 180-degrees spin-rotation on the B sites
(see Fig.~\ref{Fig:lattice}(c)) the RVB state takes the form of a translationally invariant PEPS with {\it the same} tensor on both sublattices.
Whenever open boundaries are present, one needs to assign the values of the non-contracted virtual indices at the boundaries to fully characterize 
the PEPS wave function. 

{\it Chiral RVB} --
Our goal is to find a PEPS representation of a family of states $\Psi$ with su(2),
90 degree rotation and translational
symmetry, and that contains the RVB state on a square lattice. More precisely,
let us define by $G_i$ the group transformations of $C_{4v}$ corresponding to reflection along
the vertical, horizontal, and 45 degrees axis. We want to impose that $G_i |\Psi\rangle$
coincides with $|\Psi^\ast\rangle$. For PEPS, one can enforce this
symmetry at the level of the tensor $A$ itself. This constraint, together
with the additional ones imposed by the singlet character of the PEPS fully
determine the tensor A in terms of three real parameters (see Appendix):
\begin{equation}
A=\lambda_1 R_1 +\lambda_2 R_2 + i \lambda_{\rm chiral} I  ,
\end{equation}
where the (real) $R_a$ and $I$ tensors transform according to the $B_1$ and $B_2$ 
irreducible representations (IRREP) of $C_{4v}$~\cite{landau,note_sg}, respectively. 
In other words, the $A$ tensor has $d + i\, d$ (meaning $d_{x^2-y^2} + i \, d_{xy}$) symmetry. 
On a finite $L\times L$ torus with an even number of sites, the real (imaginary) parts of the PEPS are sums of products of
even (odd) numbers of $R$ by even (odd) numbers of $I$ tensors. Therefore, the PEPS takes the form,
$
\Psi =\Psi_{s} + i\, \Psi_{g} ,
$
where the real components $\Psi_{s}$ and $\Psi_{g}$ transform according to
the $A_1$ ($s$-wave) and $A_2$ ($g$-wave) IRREP, respectively, so that
$G_i |\Psi_s\rangle=|\Psi_s\rangle$ and $G_i |\Psi_g\rangle=-|\Psi_g\rangle$. 
For $\lambda_2=\lambda_{\rm chiral}=0$, one recovers the simple nearest-neighbor RVB state~\cite{norbert,didier}. 
$\lambda_2$ and $\lambda_{\rm chiral}$ introduce longer range (AB) singlet bonds via ``teleportation''
as in the non-chiral spin liquid of Ref.~\onlinecite{wang}.
Note that we found that $|\Psi_{g}\rangle\ne 0$ only if $\lambda_2\ne 0$ and $\lambda_{\rm chiral}\ne 0$  simultaneously. 
Hereafter we fix $\lambda_1=\lambda_2=1$. 

In order to identify the above PEPS as a chiral spin liquid with topological order
 we have to prove that
(i) $\Psi$ and $\Psi^\ast$ are different in the thermodynamic
limit and (ii) there are different topological sectors on a cylinder e.g. with even or odd number of
spinons on the boundaries or with even or odd number of ($\mathbb{Z}_2$) fluxes penetrating the cylinder.
(iii) In addition, we have to fully characterize the edge physics using the PEPS bulk-boundary correspondence,
so that we can determine the Entanglement Spectrum (ES) and the Entanglement Entropy (EE).

\begin{figure}
\begin{center}
\includegraphics[width=8cm,angle=0]{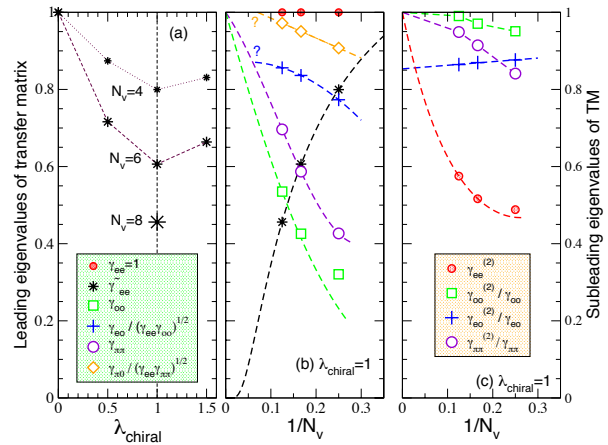}
\caption{(a) Leading eigenvalue $\tilde\gamma_{ee}$ (normalized by $\gamma_{ee}$) of the $\tilde\Gamma_{ee}$ block of the transfer matrix 
as a function of $\lambda_{\rm chiral}$ (for $\lambda_1=\lambda_2=1$) and for $N_v=4$, $6$ and $8$ (as shown on plot).
(b) Finite size scaling of the leading eigenvalues of the diagonal blocks ($\Gamma_{ee}$ and $\Gamma_{oo}$) and of the off-diagonal blocks ($\tilde\Gamma_{ee}$ and $\Gamma_{eo}$) of the transfer matrix (after proper normalization). 
The diagonal (off-diagonal) blocks have $S_z^{BK}=0$ ($S_z^{BK}=1/2$) quantum numbers.
(c) Finite size scaling of the second 
leading eigenvalues $\gamma_{\mu\nu}^{(2)}$ of the $\Gamma_{ee}$, $\Gamma_{oo}$ and $\Gamma_{eo}$ blocks (after proper normalization).
  }
\label{Fig:TM}
\end{center}
\vskip -0.5cm
\end{figure}

{\it Transfer matrix and topological sectors} -- 
To do so we place the PEPS wavefunctions $\Psi$ and $\Psi^*$ on a (horizontal) cylinder of circumference 
$N_v$ ($=4, 6, 8$) and length $N_h\rightarrow\infty$. 
Their overlap can be written as a bra-ket (BK) product
$\langle\Psi|\Psi^*\rangle = V_{\rm left} \tilde\Gamma^{N_h} V_{\rm right}$ where $\tilde\Gamma$ is a $D^{2N_v}\times D^{2N_v}$ 
transfer matrix (TM) and $V_{\rm left}$ and $V_{\rm right}$ stand 
for left and right vectors (of dimension $D^{2N_v}$) defining the boundary conditions. 
To construct $\tilde\Gamma$ one first defines a local rank-4 ${\mathbb E}$ tensor by contracting the physical index of two superposed
bra and ket $A$ tensors, namely $\tilde {\mathbb E}_{LURD}= \sum_sA^s_{lurd} A^s_{l' u' r' d'}$ where 
the indices of $\mathbb E$ combine the indices of the superposed bonds of the top (bra) and bottom (ket) $A$ tensors.
Then, one builds a (vertical) 
periodic array of $N_v$ such $\mathbb E$ tensors, contracting over the (vertical) bond indices as shown in Fig.~\ref{Fig:lattice}(d). 
Similarly, one can also construct the "regular" transfer matrix $\Gamma$ as
${\mathbb E}_{LURD}= \sum_sA^s_{lurd} (A^s_{l' u' r' d'})^*$ from which the normalization $\langle\Psi|\Psi\rangle$ can be obtained. 
If the leading eigenvalue (LE) $\tilde\gamma_{ee}$ of $\tilde\Gamma$ (normalized by the LE of $\Gamma$) 
is strictly smaller than 1, then $\Psi$ and $\Psi^\ast$ are orthogonal in the $N_h\rightarrow\infty$ limit. 
This is what is observed in Figs.~\ref{Fig:TM}(a) as soon as $\Psi$ acquires an imaginary component.
 Also, for increasing circumference, $\tilde\gamma_{ee}\rightarrow 0$ as seen in Fig.~\ref{Fig:TM}(b), implying that $\Psi$ and $\Psi^\ast$
are indeed two independent states. 

Topological properties of $\Psi$ (topological sectors, etc...) can be obtained from the ``regular''  TM~\cite{norbert2}.
It is easy to check that $\Gamma$ (like $\tilde\Gamma$) has a simple block-diagonal structure associated to a number 
of conserved quantities like the difference $S_z^{BK}=S_z^{\rm bra}-S_z^{\rm ket}$ between the $z$-components 
of the total virtual spin~\cite{note1} on the bra and the ket, namely $\Gamma=\bigoplus \Gamma [S_z^{BK}]$.
We found that the leading eigenvalues of the TM's appear in the lowest $S_z^{BK}$ sectors,
namely $S_z^{BK}=0$ and $S_z^{BK}=\pm 1/2$. Secondly, one finds that the parity of the number of virtual $|0\rangle$ 
states (or spinons at the boundaries)
is conserved {\it independently} in the bra (top) and ket (bottom) layers, as for the Kagome RVB $\mathbb{Z}_2$ liquid~\cite{norbert,didier,note2}, suggesting topological order.  The $S_z^{BK}=0$ sector of $\Gamma$ is therefore split into even-even 
(ee) and odd-odd (oo) sub-sectors, namely $\Gamma [0]=\Gamma_{ee}\oplus\Gamma_{oo}$, while the $S_z^{BK}=1/2$ sector is split into two degenerate even-odd (eo) and odd-even (oe) sub-sectors, namely 
$\Gamma [1/2]=\Gamma_{eo}\oplus\Gamma_{oe}$.  
In addition, one can insert (horizontal) strings of $\mathbb{Z}_2$ vison operators~\cite{norbert,didier} in the bra and/or ket layers of $\Gamma$ (restricting here to $\Gamma_{ee}$).
We found that the resulting TM depend only on the {\it parities} of the number of $\pi$-fluxes in each layer.
We denote by $\Gamma_{\pi0}$ ($\Gamma_{\pi\pi}$) the TM  with a vison flux in a single
(both) layer(s). 

From the above analysis we see that, a priori, one can construct four wave functions $\Psi_e^0$,
$\Psi_o^0$,  $\Psi_e^\pi$ and $\Psi_o^\pi$ characterized by the parities of the numbers of spinons at the
boundary (as specified by the subscript) and fluxes penetrating the cylinder (as specified by the superscript). However, 
we still have to check
that these wave functions, locally indistinguishable, remain truly distinct (and hence orthogonal) 
in the thermodynamic limit. 
For $N_h\gg N_v$, the overlaps 
behave as $\langle \Psi_o^0 | \Psi_e^0\rangle = (f_{eo})^{N_h}$
and $\langle \Psi_e^\pi | \Psi_e^0\rangle = (f_{\pi 0})^{N_h}$
with $f_{eo}=\gamma_{eo}/ (\gamma_{ee} \gamma_{oo})^{1/2}$ and
$f_{\pi 0}=\gamma_{\pi 0}/ (\gamma_{ee} \gamma_{\pi\pi})^{1/2}$, where 
$\gamma_{ee}$, $\gamma_{oo}$, $\gamma_{\pi\pi}$, $\gamma_{eo}$, and $\gamma_{\pi 0}$
are the LE of $\Gamma_{ee}$, $\Gamma_{oo}$, $\Gamma_{\pi\pi}$, $\Gamma_{eo}$, and $\Gamma_{\pi 0}$,
respectively.
If the wave function subspace remains degenerate in the $N_v\rightarrow\infty$ limit, we expect that (i) $\gamma_{oo}$ and $\gamma_{\pi\pi}$ converge exactly to $\gamma_{ee}$ (the largest LE set to 1) while  (ii) $f_{eo}$ 
and $f_{\pi0}$ converge to values {\it strictly smaller than 1}. 
In contrast, if $f_{\pi0}\rightarrow 1$ (or $f_{eo}\rightarrow 1$), the insertion of a flux 
in the cylinder (or a spinon at the boundaries) does not generate a new topological sector. 
The finite size scalings shown in Fig.~\ref{Fig:TM}(b) -- or in the Appendix for other parameters --
are all compatible with (i). However, extrapolations of either $f_{\pi0}$
or  $f_{eo}$ might be inaccurate in some cases -- see Fig.~\ref{Fig:TM}(b) and Appendix for other parameters -- due to slow convergence and the limitation to small perimeters.  
It is however plausible that $f_{eo}<1$ and $f_{\pi0}<1$ in the thermodynamic limit (at least for some parameters) ensuring the existence of four Indentity ($\Psi_e^0$), Spinon ($\Psi_o^0$), Vison ($\Psi_e^\pi$) and Vison/Spinon ($\Psi_o^\pi$) topological sectors (wave functions). 

{\it Asymptotic correlations} --
The behavior of the second LE $\gamma_{\mu\nu}^{(2)}$ (to be normalized by the leading one) of each block $\Gamma_{\mu\nu}$ of $\Gamma$, provides key informations on the long-distance correlations associated to the quantum number of the block
-- typically one expects $\gamma_{\mu\nu}^{(2)}/\gamma_{\mu\nu}\sim \exp{(-1/\xi_{\mu\nu})}$. 
The finite size scalings shown in Fig.~\ref{Fig:TM}(c) suggest that both $\gamma_{ee}^{(2)}\rightarrow 1$
and $\gamma_{oo}^{(2)}/\gamma_{oo}\rightarrow 1$.  We also find that the next LE are also going to $1$,
an evidence of gapless even-even and odd-odd sectors of the TM i.e. $\xi_{ee}=\xi_{oo}=\infty$. 
Correlations of spin-singlet operators are therefore expected to be algebraic (infinite correlation length).
In contrast, we find a gapped even-odd sector with an extrapolated value $\gamma_{oe}^{(2)}/\gamma_{oe}|_\infty\simeq 0.85$ 
corresponding to a magnetic correlation length $\xi_{eo}\sim 6$ (in units of the lattice spacing) for $\lambda_{\rm chiral}=1$. 

{\it Entanglement Spectrum and chiral edge modes} -- Using the standard procedure~\cite{cirac}, we have computed the ES, EE and Entanglement Hamiltonian (EH) for infinite cylinders (of circumference $N_v=4, 6$ and $8$) cut into two (semi-infinite) halves.  
The fixed-point right and left vectors, $V_R^{\mu\mu}$ and $V_L^{\mu\mu}$
obtained by applying the TM iteratively on (random) boundary vectors belonging to the $S_z^{BK}=0$ even ($\mu\equiv e$) and odd ($\mu\equiv o$) sectors are viewed as operators (of trace 1) acting on the virtual indices at the edge.
From the right and left operators $\sigma_L=\frac{1}{2} (V_L^{ee}+V_L^{oo})$ and
$\sigma_R=\frac{1}{2} (V_R^{ee}+V_R^{oo})$, one obtains the EH $H_E=-\ln{(\sqrt{(\sigma_R)^\top}\sigma_L\sqrt{(\sigma_R)^\top})}$.
$H_E$ is block-diagonal and the blocks are labelled by the momentum $K$ along the edge and by the modulus of the 
z-component of the virtual (edge) spin  $|S_z|=|S_z^{\rm bra}|=|S_z^{\rm ket}|$~\cite{note1}. 
We have computed the ES on infinite cylinders   
as a function of $K$.  
At low (quasi)energy, linearly dispersing chiral (i.e. with a definite sign of their velocity) modes with almost 
equally spaced levels are seen for all values of $N_v$ 
(see Appendix). Fig.~\ref{Fig:ES} shows the $N_v=8$ spectra which 
have been shifted and rescaled (by the same amount)
to set the average level spacing of the edge states to $\Delta=1$. 
This makes the resemblance with the spectrum of a chiral CFT very striking,
where each ``tower of states''  corresponds to a WZW primary field and 
its descendants. Since the ES inherits from the spin-singlet character of the PEPS 
SU(2) Kramer degeneracies, the $SU(2)_1$ CFT of the KL state
is the most natural candidate. 
However, the $SU(2)_1$ CFT has only two primary fields -- the Identity and the Spinon fields -- 
in disagreement with the plausible existence 
of four topological sectors for the $d+id$ RVB. In addition,  our numerical estimation of 
the conformal weight $h_{1/2}$ of the Spinon sector (from the ratio of the lowest $|S_z|=1/2$
energy level to the $|S_z|=1$ one) gives $h_{1/2}\simeq 0.93$ for $N_v=6$ and
$h_{1/2}\simeq 0.84$ for $N_v=8$, significantly above the value $h_{1/2}=1/4$ for $SU(2)_1$.

\begin{figure}
\begin{center}
\includegraphics[width=8cm,angle=0]{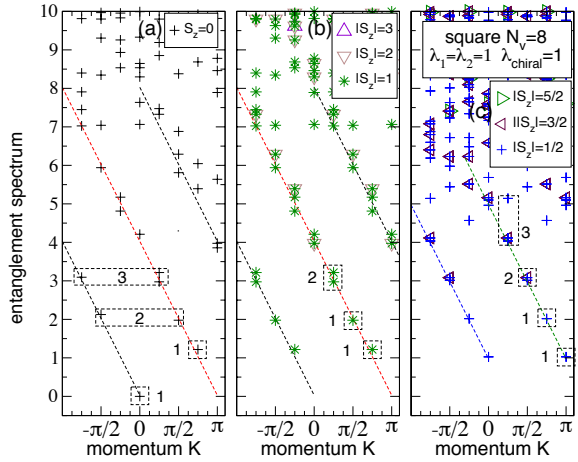}
\caption{Re-scaled Entanglement Spectrum vs momentum along the edge (for $\lambda_1=\lambda_2=\lambda_{\rm chiral}=1$).
(a), (b) and (c) show different sectors of the Z-component of the virtual spin on the edge.
The dashed lines show the chiral edge modes. 
In (a), the momentum of the states of the chiral mode is defined mod $\pi$.
The multiplicities 1, 1, 2, and 3 of the quasi-degenerate states in the boxes -- expected for a CFT -- are shown on the plot.
  }
\label{Fig:ES}
\end{center}
\vskip -0.5cm
\end{figure}

\begin{figure}
\begin{center}
\includegraphics[width=8cm,angle=0]{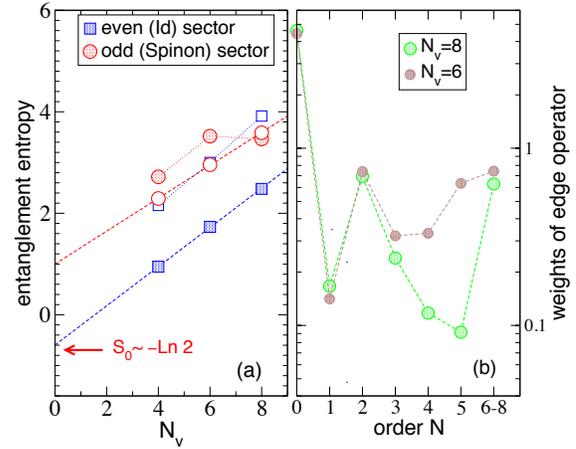}
\caption{(a) Entanglement entropy $S$ 
vs cylinder perimeter in the two topological sectors with (open symbols)
and without (full symbols) $\mathbb Z_2$ flux (for $\lambda_1=\lambda_2=\lambda_{\rm chiral}=1$). 
Linear fits (area law) $S=aN_v+b$ are shown.
(b) Weights vs $N$ of the Entanglement Hamiltonian decomposed in terms of $N$-body contributions, for $N_v=6$ 
and $N_v=8$. 
  }
\label{Fig:EH}
\end{center}
\vskip -0.5cm
\end{figure}

{\it Entanglement Entropy and Hamiltonian} -- 
We have computed the Von Neumann EE from the ES
and results for the two topological sectors (with and without $\mathbb Z_2$ flux)
are shown in Fig.~\ref{Fig:EH}(a).
In the Identity (integer spin) sector and no flux one can fit the EE according to the area law
 with a negative intersect $b\sim -0.596$. This value does not agree with the expected $-\ln{2}/2$ 
value for $SU(2)_1$ but is close to $-\ln{2}\sim 0.69$ also found in the critical RVB spin liquid~\cite{wang}. 
To investigate the range of the EH we have decomposed it in terms 
of separate $N$-body terms whose weights can be computed. 
As can be shown in Fig.~\ref{Fig:EH}(b), the BH is long-range and retains a large weight on extended
operators involving all the sites of the ring i.e. with $N=N_v$. 

{\it Discussion and outlook} -- 
It is well known in Particle Physics and lattice gauge theories (Nielsen and Ninomiya no-go theorem~\cite{NN}) that a chiral field theory cannot be discretized on a lattice. 
This has brought some doubt in the community about the possibility of constructing a PEPS (of finite bond dimension $D$)
 that could be the ground state of a local Hamiltonian within a topological chiral phase. Together with Ref.~\onlinecite{shuo},
 the present work brings new perspectives~: we have constructed a (two parameter) family of PEPS which are CSL with topological order
 and chiral edge modes and 
 share some properties with the KL spin liquid state. Nevertheless, a precise identification of the PEPS was not possible and
some differences with the KL state may arise~: (i) First, once the time reversal is broken, the KL state is
expected to have two-fold topological degeneracy on a torus. A priori, our CSL bears four topological sectors. 
The scaling of the relevant leading eigenvectors of the TM does not allow to safely conclude that two sectors only will survive in 
the thermodynamic limit.
(ii) Our numerical simulations also point towards critical (singlet) correlations. This property may be closely related to
 the bipartiteness  of the lattice and the existence of an effective field theory formulated in terms of height/gauge degrees of freedom~\cite{height}. 
 Similar construction on non-bipartite lattice like the Kagome lattice (requiring larger bond dimension $D>3$) is left for future studies. 
(iii) Lastly, a precise assignment of the edge physics to a simple CFT was not possible, also calling for further studies. 
Deviations from a simple $SU(2)_1$ CFT (expected for the KL state) may appear due to the critical 
property of the bulk. 
Finally, we note that PEPS construction of non-Abelian CSL with edge modes described by $SU(2)_k$ CFT with $k>1$ would also be of
 great interest. 

{\bf Acknowledgment} --  This project is supported by the
 NQPTP ANR-0406-01 grant (French Research Council).
The numerical computations have been achieved 
at the CALMIP UV Supercomputer (Toulouse) and valuable help
on the codes from Nicolas Renon is acknowledged.
DP\ acknowledges illuminating discussions with J\'er\^ome Dubail, Beno{\^\i}t Estienne, Pierre Pujol and Guifre Vidal
as well as with the participants of the workshop {\it ``Topological Phases and Quantum Computation''} (May 2014)
at the Moorea Ecostation Center for Advanced Studies. JIC is partially supported by the EU project SIQS. 
NS\ acknowledges support by the Alexander von Humboldt
foundation and the ERC grant WASCOSYS.

\begin{widetext}

\begin{appendices}
\section{Derivation of the PEPS ansatz.}

Here we wish to construct a modified RVB {\it complex} singlet wavefunction whose real and imaginary components 
transform according to
different irreducible representations (IRREP) ${\cal R}_1$ and 
${\cal R}_2$ of the $C_{4v}$ point group of the square lattice~\cite {landau}.  
More precisely, (i) the wavefunction should have the form $|\Psi\rangle =|\Psi_{{\cal R}_1}\rangle + i |\Psi_{{\cal R}_2}\rangle$,
where $|\Psi_{{\cal R}_1}\rangle$ and $|\Psi_{{\cal R}_2}\rangle$ are assumed to be real (and non-zero), and (ii)
${\cal R}_1$ and ${\cal R}_2$ have to be such that, under any reflection symmetry of $C_{4v}$,  
$|\Psi\rangle$ transforms into $|\Psi^\ast\rangle$, 
its time-reversed state (possibly up to a sign), which is indeed a necessary condition for a chiral spin liquid.
With PEPS, one can achieve exactly this goal by enforcing symmetries at the level of the tensor itself.
Let us write $A = R+ i I$, where $R$ and $I$ are two real tensors. We impose that  $R$ and $I$
transform, under point group operations, according to the ${\cal T}_1=B_1$ ($d_{x^2-y^2}$ orbital symmetry) and ${\cal T}_2=B_2$ ($d_{xy}$ orbital symmetry) IRREPs, respectively. In that case, it is easy to see (see main text) that the real and imaginary parts of the wavefunction (on a $L\times L$ torus) transform according to the ${\cal R}_1=A_1$ (s-wave) and 
${\cal R}_2=A_2$ (g-wave) IRREP of $C_{4v}$.

Hence, we require that
the $R$ and $I$ components of the $A$ tensor follow,
\begin{eqnarray}
R^s_{lurd}=R^s_{ldru}=R^s_{ruld}=-R^s_{drul}=-R^s_{uldr} ,  \\
I^s_{lurd}=-I^s_{ldru}=-I^s_{ruld}=I^s_{drul}=I^s_{uldr} .
\end{eqnarray}
Here $l,u,r$ and $d$ label the bond variables (0,1, and 2 for the $|\uparrow\rangle$, $|\downarrow\rangle$ and
$|0\rangle$ virtual states, respectively) clockwise around the site, starting from its left bond,
and $s=0,1$ is the physical index.
To solve these equations one has to
go through all possible combinations of the bond indices $l$, $u$, $r$ and $d$.
One obtains,
\begin{eqnarray}
R^{s'}_{2s22}&=& \lambda_1^{ss'}   ,  \\
R^{s'}_{222s}&=& \lambda_1^{ss'}  ,   \\
R^{s'}_{22s2}&=& - \lambda_1^{ss'}  ,  \\
R^{s'}_{s222}&=&- \lambda_1^{ss'}  .   
\end{eqnarray}
 In fact, spin SU(2)-invariance implies that,
\begin{equation}
\lambda_1^{ss'}= \delta_{ss'}\lambda_1 .
\end{equation}
These tensor elements correspond exactly to those of the NN RVB wavefunction.  
Note that the tensor elements of $I$ with the same indices identically vanish by symmetry.
The solution of Eqs. (2) and (3) gives also non-zero tensor elements that correspond to quantum teleportation via diagonal bonds,
\begin{eqnarray}
R^{s'}_{s{\bar s}s2}&=& \lambda_0^{ss'}   ,  \\
R^{s'}_{s2s{\bar s}}&=& \lambda_0 ^{ss'}  ,  \\
R^{s'}_{2s{\bar s}s}&=& - \lambda_0^{ss'}   ,  \\
R^{s'}_{{\bar s}s2s}&=& - \lambda_0^{ss'}   ,  
\end{eqnarray}
where $\lambda_0^{ss'}\in {\mathbb R}$. One also gets (grouping the $R$ and $I$ tensors),
\begin{eqnarray}
A_{ss{\bar s}2}^{s'}&=& \lambda_2^{ss'}  + i \lambda_{\rm chiral}^{ss'}  ,  \\
A_{{\bar s}2ss}^{s'}&=& \lambda_2^{ss'}  + i \lambda_{\rm chiral}^{ss'}  ,  \\
A_{s2{\bar s}s}^{s'}&=& \lambda_2^{ss'}  - i \lambda_{\rm chiral}^{ss'}  ,  \\
A_{{\bar s}ss2}^{s'}&=& \lambda_2^{ss'}  - i \lambda_{\rm chiral}^{ss'}  ,  \\
A_{ss2{\bar s}}^{s'}&=& -\lambda_2^{ss'}  + i \lambda_{\rm chiral}^{ss'}  ,  \\
A_{2{\bar s}ss}^{s'}&=& -\lambda_2^{ss'}  + i \lambda_{\rm chiral}^{ss'}  ,  \\
A_{s{\bar s}2s}^{s'}&=& -\lambda_2^{ss'}  - i \lambda_{\rm chiral}^{ss'}  ,  \\
A_{2ss{\bar s}}^{s'}&=& -\lambda_2^{ss'}  - i \lambda_{\rm chiral}^{ss'}  ,  
\end{eqnarray}
where  $\lambda_2^{ss'}, \lambda_{\rm chiral}^{ss'}\in {\mathbb R}$ are independent constants
and $\bar s$ is the time-reversed of the spin $s$.

Finally, enforcing  the invariance under SU(2) spin-rotations (singlet character of the PEPS) one finds that
$\lambda_2^{ss'}$ ($\lambda_{\rm chiral}^{ss'}$ and $\lambda_0^{ss'}$) is (are) diagonal in the spin indices and
even (odd) under spin inversion,
 \begin{eqnarray}
\lambda_2^{ss'}&=&\delta_{ss'} \lambda_2, \\
\lambda_{\rm chiral}^{ss'}&=&\delta_{ss'} (-1)^s \lambda_{\rm chiral}, \\
\lambda_0^{ss'}&=& \delta_{ss'} (-1)^s\lambda_0.
\end{eqnarray}
Note that spin rotation invariance leads also to the relation,
\begin{equation}
\lambda_0=2\lambda_2.
\end{equation}

\section{Leading TM eigenvalues: comparison between different wave functions.}

\begin{figure}
\begin{center}
\includegraphics[width=8cm,angle=0]{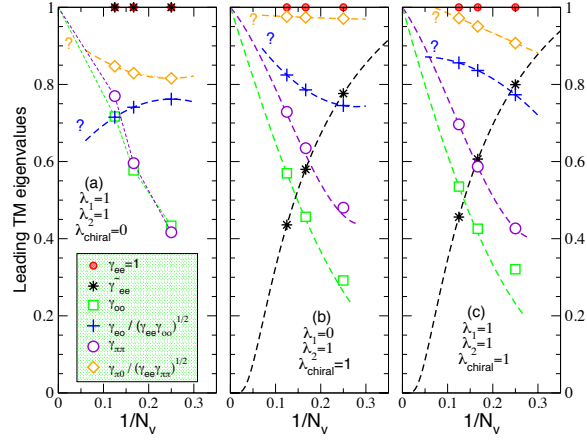}
\caption{Finite size scaling of the leading eigenvalues of the diagonal blocks ($\Gamma_{ee}$ and $\Gamma_{oo}$) and of the off-diagonal blocks ($\tilde\Gamma_{ee}$ and $\Gamma_{eo}$) of the transfer matrix (after proper normalization). 
Same notation as Fig.~\ref{Fig:TM}(b) in the main text. The three panels correspond to different choices 
of the parameters of the PEPS as indicated 
on the plot. (a) is a non-chiral RVB state ($\tilde\gamma_{ee}=1$), (b) does not contain NN valence bonds and (c) is a repetition of
 Fig.~\ref{Fig:TM}(b) for convenience.  }
\label{Fig:TM2}
\end{center}
\vskip -0.5cm
\end{figure}

We compare the finite size scalings of the LE of the TM for three choices of the tensor parameters in Fig.~\ref{Fig:TM2} corresponding to the critical (non-chiral) RVB state in (a)
and chiral topological liquids in (b) and (c). 
For all wave functions, the finite size scalings are consistent
with $\gamma_{oo}\rightarrow 1$ and $\gamma_{\pi\pi}\rightarrow 1$.
However, in (b) and (c) it is difficult to assess with full confidence that 
$\gamma_{\pi 0}/ (\gamma_{ee} \gamma_{\pi\pi})^{1/2}$ and $\gamma_{eo}/ (\gamma_{ee} \gamma_{oo})^{1/2}$ {\it do not} go to exactly $1$ when $N_v\rightarrow\infty$, as required to obtain
separate topological sectors. 

\section{Entanglement Spectrum: comparison between different sizes.}

Comparisons between ES for cylinders with different perimeters 
is shown in Fig.~\ref{Fig:ES2}(a-c). Remarkably the slope of the chiral edge modes is
very similar for $N_v=6$ and $N_v=8$. 

\begin{figure}
\begin{center}
\includegraphics[width=8cm,angle=0]{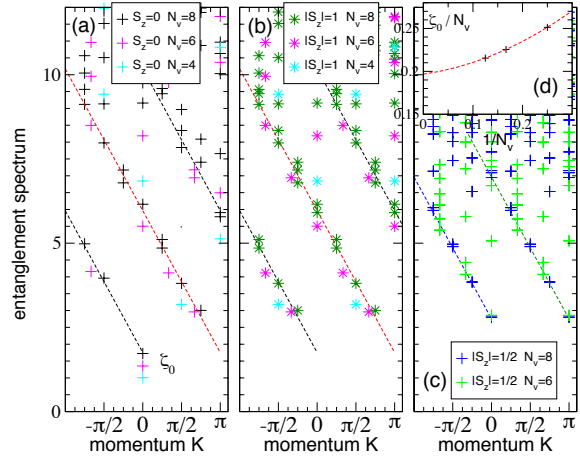}
\caption{Entanglement Spectrum vs momentum along the edge for $N_v=4, 6, 8$
(for $\lambda_1=\lambda_2=\lambda_{\rm chiral}=1$).
(a), (b) and (c) show $S_z=0$, $|S_z|=1$ and $|S_z|=1/2$, respectively.
The dashed lines show a fit of the chiral edge modes for $N_v=8$. (d) Finite size scaling of the lowest eigenvalue
of the ES.  }
\label{Fig:ES2}
\end{center}
\vskip -0.5cm
\end{figure}

\end{appendices}
\end{widetext}


\begin{thebibliography}{99}
\bibitem{note_sg} One could equivalently re-write the same PEPS using new $R_a$ and $I$ tensors belonging
to the $A_1$ (s-wave) and $A_2$ (g-wave) IRREP, respectively, via a simple gauge transformation.
\bibitem{note2} For the NN RVB state on the square lattice, the number of $|0\rangle$
itself is conserved. It is a consequence of the existence of longer-range singlets (when $\lambda_2\ne 0$ and/or $\lambda_{\rm chiral}\ne 0$) that only the parity is
conserved.
\end{thebibliography}
\end{document}